\documentclass[fleqn,10pt]{SelfArx} 

\usepackage[english]{babel} 

\usepackage{fancyvrb}
\usepackage{xcolor}

\definecolor{color1}{RGB}{0,0,90} 
\definecolor{color2}{RGB}{0,20,20} 

\usepackage{hyperref} 
\hypersetup{hidelinks,colorlinks,breaklinks=true,urlcolor=color2,citecolor=color1,linkcolor=color1,bookmarksopen=false,pdftitle={Title},pdfauthor={Author}}

\JournalInfo{Arxiv} 
\Archive{Computing Research Repository} 

\PaperTitle{LUCE: A Blockchain Solution for monitoring data License accoUntability and CompliancE} 

\Authors{Andine Havelange\textsuperscript{1}, Michel Dumontier\textsuperscript{1}, Birgit Wouters \textsuperscript{3}, Jona Linde \textsuperscript{2}, David Townend \textsuperscript{3}, and Arno Riedl \textsuperscript{4} and Visara Urovi\textsuperscript{1}} 
\affiliation{
\textsuperscript{1}\textit{Institute of Data Science (IDS), Maastricht University, The Netherlands} \\
 \textsuperscript{2}\textit{School of Business and Economics (SBE), Maastricht University, The Netherlands} 
\\ 
\textsuperscript{3}\textit{Health Ethics and Society (HES), Maastricht University, The Netherlands}
\\ 
\textsuperscript{4}\textit{CESifo, IZA, Netspar and Department of  Microeconomics and Public Economics, Maastricht University, The Netherlands} 
} 


\begin{document}


\maketitle 


\section*{Abstract}

In this paper we present our preliminary work on monitoring data License accoUntability and CompliancE (LUCE). LUCE is a blockchain platform solution designed to stimulate data sharing and reuse, by facilitating compliance with licensing terms. The platform enables data accountability by recording the use of data and their purpose on a blockchain-supported platform: LUCE allows for individual data to be rectified and erased. In doing so LUCE can ensure subjects' General Data Protection Regulation's (GDPR) rights to access, rectification and erasure. Our contribution is to provide a distributed solution for the automatic management of data accountability and their license terms.

\section{Introduction}
 
There is an increasing awareness in the scientific community about the importance of data sharing  \cite{OpenData2017, Nelson2009}. As a key element of scientific research, data sharing allows reproduction of scientific results \cite{Borgman2012, Nelson2009, OpenData2017} and helps prevent data fabrication and falsification \cite{Tenopir2011}. Indeed, data sharing is now often required by funding bodies, publishers \cite{Borgman2012, OpenData2017} and several EU and US funding initiatives \cite{Elixir2017,EOSC2018, NCATS2018,DataCommons2018}.

\noindent Although the benefits of data sharing and reuse are widely acknowledged, evidence of data sharing practices is limited. Currently, a majority of researchers share their data only directly, that is, from person to person (e.g. by email), and mostly with collaborators, which suggests that trust is an important factor in sharing data. One third of researchers do not share their data at all  and public data sharing occurs through the appendix of research articles, stand-alone publications in data journals, repository, personal websites and specific websites \cite{OpenData2017,Borgman2012}. These ways of data sharing make it hard to find data even if it is shared publicly.

\noindent Several factors explain the large gap between the agreed importance of sharing data and today's actual practices \cite{OpenData2017, Borgman2012, WhitePaper2018, Nelson2009}: (i) There is no commonly accepted definition of what data sharing exactly means and which data should be shared. (ii) Researchers lack expertise, training, infrastructure and resources, to share their data. (iii) Researchers rarely receive credit for sharing data, partly because how to cite and attribute data is not commonly defined yet. (iv) Concerns regarding privacy, control over what happens to the data and ethical issues prevent researchers from sharing their data.

\noindent One way to deal with these concerns is to use data licensing \cite{Stodden2009}. Licenses clearly state what can and what cannot be done with the data, thereby creating legal clarity for the researchers who reuse the data. Yet, a significant proportion of shared data is not licensed, which is mostly due to a lack of awareness of the need for data licensing from researchers. 
Even when data are released under a license, several issues arise. Firstly, there exist a variety of licenses to chose from \cite{RDP}. Secondly, researchers reusing data may not comply with the licensing terms because they do not read the legal text, or because they do not understand certain legal terms, or simply because they forget about the licensing terms while reusing the data (e.g. share-alike). This is further complicated by the lack of enforceability of licensing terms in the context of data sharing in scientific research. To date, it is still unclear how to facilitate understanding and conformance to one or more licensing terms in the context of data reuse.

\noindent The difficulty of tracking how data are used once they have been shared, makes those scientists sharing data to choose more restrictive licenses \cite{OpenData2017}. Moreover, when data contains information about data subjects, being able to track the use of data is even more crucial, for example in the context of the EU General Data Protection Regulation (GDPR) \cite{GDPR}. This regulation includes, among others, the right to access for individuals. This describes access rights to information pertaining individual data, such as who is using the data and for what purpose. Two other elements of the GDPR are of high significance in the context of data sharing and reuse: the individual's rights to erasure (involving the deletion of individual records) and to rectification (involving the update of individual records).

\noindent The aim of this paper is to enhance data sharing and reuse by facilitating compliance with data licensing terms for researchers, and by devising a technological framework helping researchers comply with the GDPR's rights to access, rectification and erasure. In this context, our contribution is to provide a blockchain-based solution for the automatic management of data licensing terms and for data accountability. Our solution enables data accountability in that it allows recording the purpose and the use of data, as well it enables the actors to rectify and erase data upon data subject's request. 
  
\noindent The remainder of this paper is organized as follows. In the Background section, we provide information about the building blocks of our solution, namely blockchain technology, smart contracts, the Creative Commons licenses and the EU General Data Protection Regulation. We then present related work pertaining to data licensing and data accountability. In the subsequent section, we then provide a general overview of our solution. The LUCE architecture is then explained in detail in the Architecture section. Thereafter we present a first prototype of the smart contract used in LUCE and conclude  this paper with a discussion on current and future works.

\section{Background}
In this section, we present background information about the main building blocks of our solution. We first explain blockchain technology, as well as smart contracts and Ethereum. Then, the Creative Commons Rights Licenses are introduced. Finally, the EU General Data Protection Regulation, and more specifically its key elements for data sharing and reuse, are exposed.   
\subsection{Blockchain technology}
Our proposed solution is based on blockchain technology. \\
Blockchains are distributed and immutable ledgers of transactions \cite{nakamoto2008bitcoin, Blockhub2017, Pilkington2016}. Anyone who is part of the blockchain network can inspect past transactions and add new ones to the ledger, but no one can modify them. The blockchain network is distributed. Participants hold a copy of the ledger and can add transactions to the ledger by validating them. As such, there is no central entity controlling it, thereby removing the need for a common trusted third party. 
\\ 
Blockchain can be thought of as an append-only and distributed transactional database. The blockchain network has three main components: (i) The blockchain itself, that is, the file containing the records of all transactions. (ii) The peer-to-peer (P2P) network where the participants interact via the blockchain protocol to transact and update the blockchain. (iii) The consensus mechanism, also called blockchain protocol. The blockchain stores the records of all transactions that happened since the creation of the blockchain network. The transactions are organized in blocks that are linked to each other in a chronological order. Each block contains several records of transactions and the identifier of the block preceding it in the chain \cite{nakamoto2008bitcoin}. The blocks are linked and secured using cryptographic techniques. The records of transactions that are stored in the blockchain are auditable and verifiable but cannot be modified once they have been added. Given that there is no single central authority managing the blockchain, a consensus mechanism is necessary to formally encode rules regarding how transactions are validated and how they are added to the ledger. Nodes in the P2P network use the consensus protocol to validate blockchain transactions.
\\
The blockchain protocol defines a set of consensus rules that formalize how the stakeholders in the network interact with each other. The rules describe how, and on which conditions, transactions are validated and how a block is added to the blockchain \cite{nakamoto2008bitcoin, ethereum}. When a transaction has been generated, it is broadcasted to the peer-to-peer network where it propagates from node to node \cite{Blockhub2017, edurekaBlockchain1}. Prior to transmitting the transaction to its neighbours, a node first verifies the transaction (this is possible thanks to the fact that each transaction has a unique digital signature proving that it is genuine) so that only valid transactions will be propagated through the network. Adding valid transactions to a new block and then appending that block to the blockchain is the performed by specific nodes, called the miners \cite{edurekaBlockchain1}. Like other nodes, miners validate transactions and propagate them to their neighbours; in addition, they aggregate transactions into a candidate block and add it to the blockchain once it is complete (blocks have a limited size). Depending on the model of blockchain, the miners may receive a reward in the form of tokens (e.g. Bitcoins for the Bitcoin-blockchain and Ether for the Ethereum-blockchain \cite{ethereum}). There exist various mechanisms for validating candidate blocks and adding them to the blockchain. The two best known ones are proof-of-work and proof-of-stake. 
\begin{itemize}
\item In proof-of-work, the miners have to solve a cryptographic puzzle associated to the block being created in order to add it to the blockchain \cite{Pilkington2016}. Solving this puzzle is extremely costly in terms of energy and computational power. Several miners compete at the same time to add their candidate block to the blockchain. The first one that solves this puzzle broadcasts their candidate block together with the solution of the puzzle to the peer-to-peer network. 
\item In proof-of-stake the probability of adding a block to the blockchain depends on the stake of the miners \cite{Blockhub2017} in the blockchain network. In other words, the miner who will add the next block to the blockchain is chosen in a probabilistic way, based on the amount of currency they own, also called their stake. The higher the stake, the larger the probability that a node can validate a block and add it to the blockchain.
\item Other consensus mechanisms exist, such as delegated proof-of-stake, proof-of-burn, proof-of-authority and so forth (see \cite{Blockhub2017} for more details). 
\end{itemize}
There also exist various kinds of blockchain, defined by the extent to which they are public. Three main categories can be distinguished \cite{Pilkington2016}.
\begin{itemize}
\item In a public blockchain, everyone can, freely and unconditionally, join the network and participate in determining which blocks are added to the blockchain. This also implies that everyone can inspect transactions that are stored in the blockchain and add new ones. The consensus mechanism used in this kind of blockchain is generally proof-of-work or proof-of-stake.
\item Private blockchains are managed by a single organization which grants authorization or not for adding transactions to the blockchain. Inspection-rights are either public or restricted and identities of the stakeholders are known. The use of the term blockchain in this case is subject to debate given that it is not decentralized at all. 
\item There exists a continuum of blockchains between public and private ones. Consortium or federated blockchains constitute a hybrid solution. Like private blockchains, the right to participate in the blockchain is restricted and the inspect-permission is either public or permissioned. However, they are partially decentralized in that they are managed by a group of organizations. 
\end{itemize}
The properties offered by blockchain networks have the needed flexibility to share and monitor data in a distributed leaderless manner. In this way, the specific “nodes” holding pieces of datasets of information can safeguard data ownership. There is no need to centralize data or to centralize the decision on what can be done with the data as the distributed blockchain network can agree on when and if data can be shared with other nodes. Moreover, the auditability of the network means that it can also be find out what happens to the data, who accessed them and for what purpose.


\subsection{Smart contracts and Ethereum}

The second building block needed to fully benefit from the advantage of using a blockchain network for data sharing is the notion of smart contract. A smart contract is a computer code that runs on the top of a blockchain network and that contains rules - rights and obligations - defining the interaction between the parties to the smart contract \cite{smartContract}. When all parties meet pre-defined conditions, the smart contract automatically enforces the agreement between them. Transactions with the smart contracts are recorded within the blockchain on the top of which it is running. Smart contracts allow for several parties who do not especially trust each other to transact with each other without a trusted third party, thereby also reducing transaction costs. Smart contracts can be used for simple economic transactions but also for other, more complex, purposes, like registering ownership or intellectual rights. 
\\
LUCE is implemented on top of the Ethereum Blockchain platform \cite{ethereum}, currently the most widely used platform for using smart contracts. Ethereum is an open-source and public blockchain-based distributed computing platform for building decentralized applications. 
Choosing Ethereum implies that LUCE’s blockchain can be public, meaning that anyone can inspect the transactions as well as add and verify new ones, including people who are not academics. 

\subsection{Creative Commons Rights Expression Language}
The Creative Commons \cite{CC} is an initiative that produces free and easy-to-use licenses to help people share their work in a standardized way but on the conditions of their choice. \\
\indent The Creative Commons (or CC) licenses have a three layer structure. The first one is the legal text of the license and is called the Legal Code layer of the license. However, since not only lawyers but also common people, such as researchers or educators, use the CC licenses, the second layer is the Common Deed, or human-readable, layer. It summarizes the key terms and conditions of the license. The third layer of the CC licenses is the machine-readable one. It also summarizes the key terms and conditions of the license but in a language that is understandable by software, search engines and other types of technologies. Note that only the Legal Code layer is the license, the other two layers are only a summary of it intended to make it easier for humans and computers to understand the license. As such, only the Legal Code is legally binding. 
\\
\indent The Creative Commons developed a standardized language to describe licensing terms that are machine interpretable, the Creative Commons Rights Expression Language (ccREL)\cite{CCREL2008}. ccREL is based on Resource Description Framework (RDF), which is a framework used to describe entities on the web using URL's and URI's. RDF descriptions are triples. They include the subject (e.g. a dataset), the property that is attached to that subject (e.g. a license is attached to the dataset) and the value of that property (e.g. that license in question is the Creative Commons CC-BY-NC 4.0 license). These three elements are expressed as URL's or URI's.
\\
\indent ccREL is specified as an extensible set of properties that can be associated with a licensed object \cite{CCREL2008}. More specifically, ccREL distinguishes two classes of properties, namely work and license properties. The work properties describe the licensed work and include its title, the name and the URL to cite when giving attribution, the type of work, the original source of a modified work and the URL giving additional permission beyond those in the license. The license properties include which types of use of the work are permitted, which ones are prohibited, what is required when using the work, the corresponding Legal Code of the license, possibly the legal jurisdiction to which the work is associated and possibly the data on which the license has been deprecated. The values that the \textit{permission} property can take are reproduction, distribution and derivation of the work. What might possibly be prohibited is using the work for commercial purpose and what can be required when using it is noticing which license is attached to it, attribution, share alike (e.g. using the same license in case derivative work is redistributed) and providing the source code. For instance, the Creative Commons CC-BY-NC 4.0 license permits derivation, reproduction and distribution, prohibits commercial use and requires attribution and noticing that this license is attached to the work.\\
\indent ccREL is specified in an abstract syntax-free way. This means that it constitutes a vocabulary of properties and their values but that RDF triples can be expressed independently of any particular syntax. The Creative Commons nonetheless recommend some specific concrete default syntaxes, such as RDFa for HTML pages and XMP for free-floating content such as media files. In the context of this project, it would be worth investigating the use of the Open Digital Rights Language (ODRL) as syntax for ccREL because of its extensibility. ODRL is a policy expression language that provides a flexible and interoperable information model, vocabulary, and encoding mechanisms for representing statements about the usage of content and services \cite{ODRL}. This is actually already done by Licentia \cite{Licentia}, which is a licensing assistant. Through this interface, people can choose under which conditions they want to share their work and Licentia suggests several compatible standard licenses based on the person's choice. It also allows to visualize and download standard licenses expressed using ODRL. 
\begin{figure*}[!h] 
    \centering
    \includegraphics[clip=true, scale = 0.40]{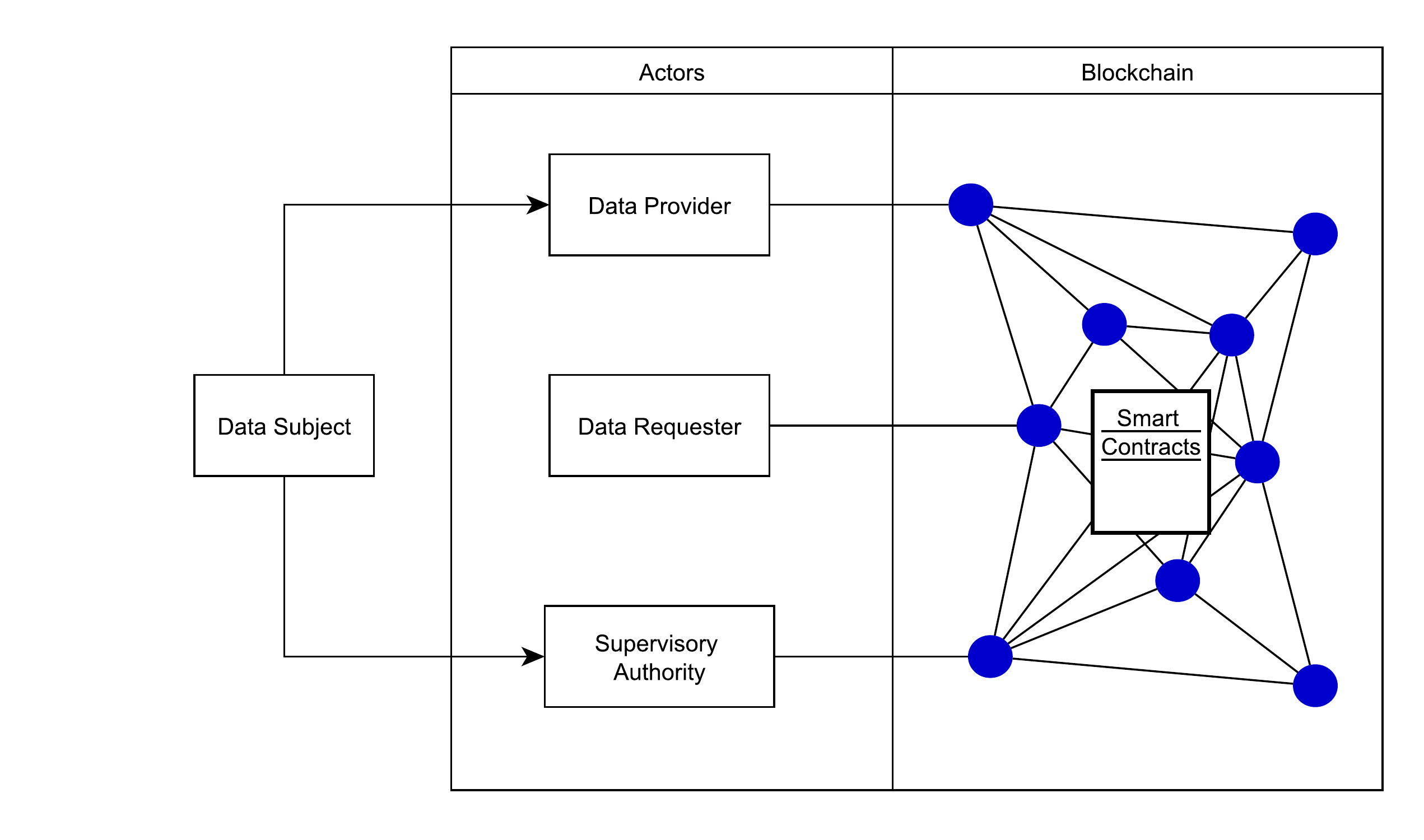}
    \caption{Connecting the data-sharing actors via blockchain}
    \label{BigPic}
\end{figure*}
\subsection{GDPR}
The new EU General Data Protection Regulation (GDPR) came into effect in May 2018 to extend the requirements of organizations in regards of collecting and processing personal data of EU residents \cite{GDPR}. Being a regulation, its rules are to be enforced in all EU member states. The GDPR also applies to non-EU organizations that collect and process EU residents’ data. 
The main actors involved in the GDPR are the following: a \textit{data subject} who is an identified or identifiable natural person whose data are contained in the dataset; Data subjects can authorize a \textit{data controller} (which can be an organisation) to access his or her personal data, with the possibility to transfer these to a \textit{data processor} (which can also be an organisation) in charge of processing these data. Another key actor is the \textit{Supervisory Authority} which is a controlling body. Each EU member state has its own National Supervisory Authority. In the context of scientific data sharing and reuse, individuals whose data records are collected by a researcher, are the data subjects and the researcher can be seen as a data controller. In case the researcher shares the dataset constituted by all the subjects’ data records and that another researcher reuses it, the second researcher can be seen as a data processor. The Supervisory Authority remains unchanged. 
Three articles of the GDPR are of crucial importance in the context of data sharing and reuse: the rights of access by the data subject (article 15), the right to rectification (article 16) and the right to erasure (article 17). The right of access states that the data subject has the “right to obtain from the controller confirmation as to whether or not personal data concerning him or her are being processed” \cite{GDPR}. If that is the case, the controller also has to give the data subject access to her or his personal data, as well as some additional information such as the purpose of the processing and to whom their data has been transferred. The right to rectification states that the data subject has the right to have his or her personal data rectified by the controller, in case they are inaccurate. Finally, if the data subject exercises his or her right to erasure, the controller has the obligation to erase the subject's personal data.  
A fourth element of crucial importance, in the case of data sharing and reuse, is that of consent. Indeed, the GDPR requires that the data subjects freely give their informed consent to the collecting and processing of their data. Importantly, under preamble 33 a broader consent for scientific research is possible, however the main legal basis of consent for secondary use are defined within article 6, paragraph 4 of the GDPR document.  
Under this article, if personal data are shared and reused, it is vital that the purpose for which they are reused is compatible with the original purpose for which they were collected. Moreover, the data provider should also decide, inter alia, on the link for which the data was collected and the intended secondary use, the context for which the data were collected, the nature of the data, consequences of further processing to data subjects and the existence or not of safety guards. As the primary data collector, the data provider, remains responsible for the use of data and the main contact point for the data subjects. 
Given that its goal is to enhance data sharing and reuse, LUCE supports dealing with these constraints so that researchers are enabled to comply with GDPR when sharing and reusing data. Nonetheless, LUCE also can support sharing of datasets that do not contain personal information, such as geological or meteorological data. 

\section{Related works}


Standard and easy-to-use licenses produced by organizations such that researchers do not have to write their own licenses already exist. A well-known example is the above-mentioned Creative Commons organization \cite{CC}. Another similar initiative is the Open Data Commons \cite{ODC}, which provides licenses for open data. There also exist some initiatives aiming at guiding researchers through the licensing landscape, such as Licentia \cite{Licentia} or the European Data Portal's licensing assistant \cite{EDP}. Another related initiative is the set of criteria developed by the (Re)Usable Data Project \cite{RDP} to measure how well the data to which the license is attached can be reused. Examples of such criteria are the existence of a standard license, or the absence of terms restricting the reuse of the data. All these initiatives facilitate researchers to license their data and help them understand licensing terms when reusing data. However, they do not allow for the enforcement of the licensing terms, and the lack of such enforcement clearly hinders data sharing. 
\\
Digital Rights Management (DRM) technology allows for such enforcement by automatically controlling access and actions that can be performed on them, by humans or computers \cite{arnab2004digital, zeng2011multimedia}. Current DRM systems manage rights related to copying, to where usage is permitted (only one or several devices), to how many times a content can be used and by whom, and so forth. However, DRM technologies are designed for digital media such as audio and video files, texts and pictures. Given that the ways and purposes of reuse are different for data than for pictures or video files, DRM technologies seem to be ill suited to deal with some common situations in data reuse, like data integration (combining several datasets) and derivative work. Additionally, DRM technologies do not allow for researchers to comply with the GDPR rights to access, rectification and erasure, which is necessary when they share data pertaining to individuals. 
\\
Data provenance are the meta-data describing the history of data, such as where they originate from, their owner and the changes made to them, as well as by whom these changes were made \cite{Ramachandran2017,Liang2017}. DataProv \cite{Ramachandran2017} and ProvChain \cite{Liang2017} are both blockchain-based solution for data provenance accountability. Once collected, data provenance records are published to the blockchain network, verified and eventually added to the blockchain. In other words, DataProv and ProvChain provide an immutable and secured ledger for data provenance records. Both solutions deal with data stored in the cloud. Even though these solutions seem to provide a way for researchers to comply with the GDPR’s right to access, how to comply with the rights to rectification and erasure is not covered. Furthermore, the compliance with licensing terms of shared data is not taken into account. 
\\
Neisse et al. \cite{Neisse2017} also propose a blockchain-based solution for data provenance accountability. However, instead of dealing with cloud data, it focuses on individuals’ (or data subjects’) data. Their solution aims at empowering data subjects by enabling them to track who has accessed their data and whether these were used accordingly to their consent. It also aims at helping data controllers prove that they actually have received that consent, the proof of which is a smart contract to which both the data controller and subject are parties. The smart contract also encodes the policies for data access, usage and transfer, as well as data provenance information. This solution focuses on the relation between data subjects and data controllers while the goal of LUCE is to focus on the relation between data controllers (researchers sharing a dataset) and data processors (researchers reusing a dataset), while supporting the data subjects into excessing their rights. The advantage of our approach is that collected datasets can be shared and reused by data processors while maintaining control on both the type of reuse and on who has access to the dataset.  
\\
The Ocean Protocol \cite{Ocean} is an industry wide initiative to implement market places for data sharing. This blockchain-based solution aims at facilitating the sharing of datasets (as well as algorithms and services, such as storage and processing) in a transparent, traceable and trustworthy way, with the data providers keeping control over their datasets. However, the Ocean Protocol mainly targets companies who collect data. Indeed, data providers can choose to be monetarily rewarded for sharing their data. It therefore seems ill suited for data sharing between researchers in the context of scientific research. Furthermore, even though the Ocean Protocol allows for traceability regarding what happens with datasets, it does not allow for recording the purpose for which the data have been used, thereby making it hard for researchers to comply with the GDPR’s right to access. Nor does the Ocean Protocol enable researchers to comply with the GDPR’s rights to rectification and erasure in an easy manner, as the compliance with these rights is not actively taken into consideration. 
\\
LUCE aims at filling the above-identified gaps by providing a solution which allows researchers to share and reuse data while enforcing data licensing terms, as well as complying with the GDPR’ rights to access, rectification and erasure. Our solution deals with full datasets (as opposed to individuals’ data records) and it mainly focuses on researchers in academia.

\section{LUCE: A solution to license accountability and compliance}
\begin{figure*}[!h] 
    \centerline{
    \includegraphics[clip=true, scale = 0.5]{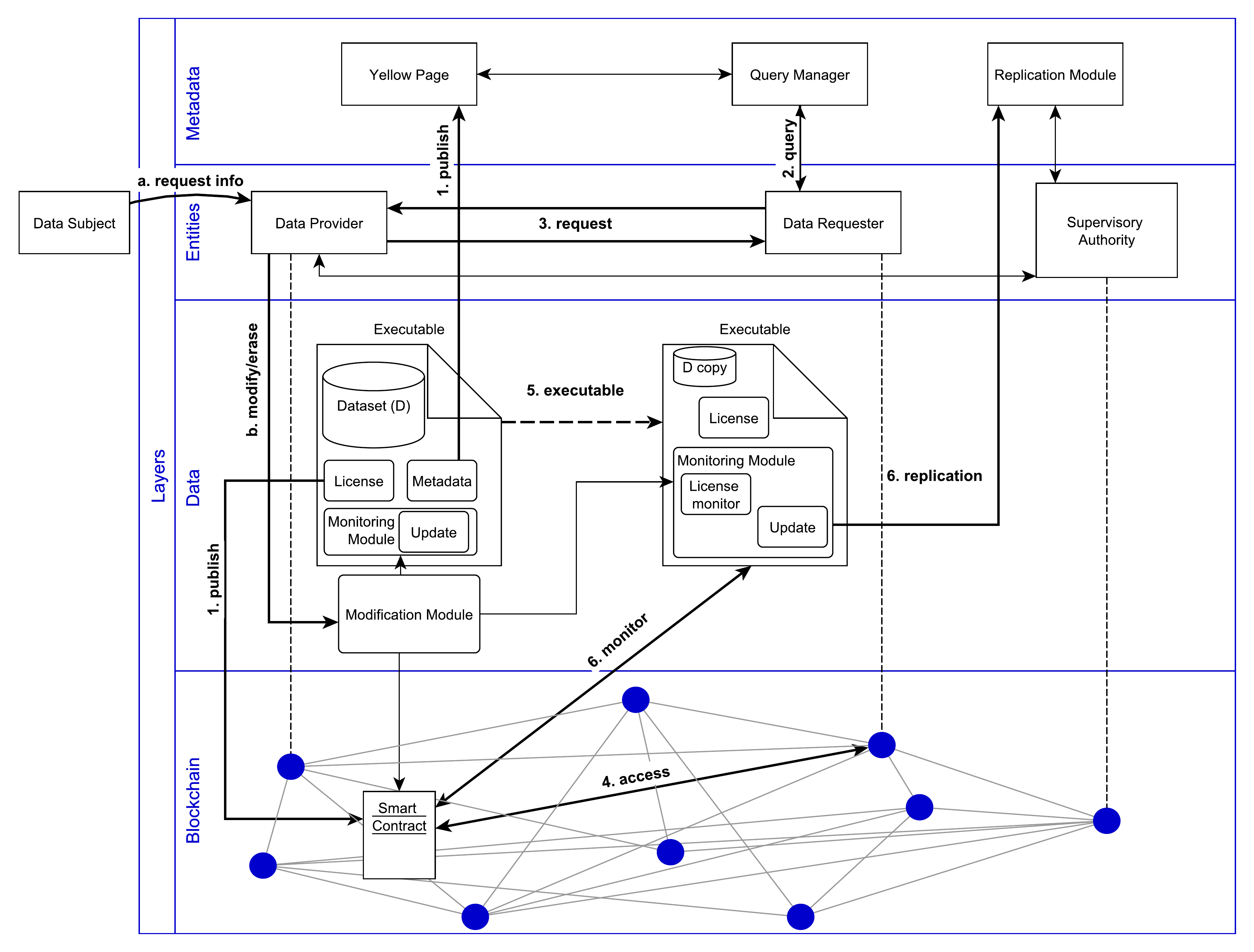}}
    \caption{LUCE Architecture}
    \label{Archi}
\end{figure*}
We propose a model which i) automatically manages and enforces licensing terms attached to a dataset, ii) records and makes available information pertaining to how and for which purpose the dataset is reused and, iii) enables compliance with the GDPR's rights to access, rectification and erasure. 
Figure \ref{BigPic} shows the main actors involved in data sharing and reuse, as well as how they interact with our solution. There are three main actors involved: the \textit{data provider}, e.g. a researcher willing to share a dataset, the \textit{data requester}, e.g. a researcher requesting to reuse a dataset, and the \textit{supervisory authority}, e.g. a national public authority in charge of monitoring the adherence to the GDPR. Although \textit{data subjects}, e.g. any individual whose data are being collected, held or processed, is are currently not directly interacting with our solution, we still take them into account as they can trigger interactions between our solution and data providers or supervisory authorities, either by exercising their rights to access, erasure or modification, or by lodging a complaint via the supervisory authority. 
\\ 
One smart contract is generated per dataset and it is used for two purposes. Firstly, each smart contract stores information that could be required by data subjects when exercising their rights. Secondly, the smart contract manages the interactions between the shared datasets and the data requesters. For instance, if a data requester stops complying with the licensing terms of a dataset, the smart contract does not grant him or her access to the dataset any more.
The smart contracts run on top of a blockchain network which is used to store information related to the interactions between the data requesters and the smart contracts. For instance, the fact that the data requester agrees with the license attached to the dataset they want to reuse, is considered as a transaction with the smart contract associated with that dataset and is stored in the blockchain. Each data provider, data requester and supervisory authority that is using our solution is a node in the peer-to-peer blockchain network.\\ 
Another component of our solution is checking that the data requester is complying with the license when using the dataset and periodically reporting to the smart contract. Data providers using our solution must thus choose a license that is expressed in a machine-readable language. In our work we use the Creative Commons Expression Language (ccRELl) for monitoring licensing terms.



\section{Architecture}

LUCE focuses on three main scenarios: \textit{sharing a dataset, reusing a dataset} and \textit{complying with the GDPR’s rights} to access, rectification and erasure. We make some assumptions regarding these scenarios:
\begin{enumerate}
\item Datasets are shared as a whole. This means that a data requester cannot ask only 
 a portion of the dataset.
\item Dataset integration is not taken into account, meaning that we currently do not cover cases in which several datasets are combined when reused. 
\item All the records of data subjects in a dataset have given the same consent to the data provider who is sharing the data.
\item The license under which the dataset is shared is compatible with the consent of the data subjects. 
\item The data records in a shared dataset are anonymized and only the data provider can map a data subject to their records in the shared dataset. 
\end{enumerate}
Assumptions 3-5 are only relevant in case the shared dataset contains data records of data subjects. Assumptions 2, 3 and 4 will be removed in future developments of LUCE. 
A detailed overview of LUCE’s architecture is displayed in figure \ref{Archi}. In what follows, we will explain how LUCE deals with the three aforementioned scenarios. 
To increase the efficiency of data discovery and access, semantic descriptions of the datasets in the form of meta-data are needed. We take the ADA-M profile \cite{woolley2018responsible} approach as our starting point for both describing the data and then requesting access to them. ADA-M profiles define an information model for producing structured meta-data “Profiles” of regulatory conditions, thereby enabling efficient application of those conditions across regulatory spheres. In brief, ADA-M includes: a) A header section for contextual information about the ADA-M Profile itself and some basic statements about the data; b) A body section for specifying regulatory concepts into “Permissions” (mainly relevant to consent), “Terms” (typically relating to legal/contractual matters), and “Meta-Conditions” (over-arching topics) (see \cite{woolley2018responsible} for a detailed description). 

\subsection{Sharing a dataset}
The first scenario covered by LUCE is a data provider sharing a dataset. . 
\paragraph{Publish}The data provider shares a dataset by publishing its meta-data in the \textit{Yellow Page}, an on line shared directory where data requesters, can search for shared datasets. The shared meta-data include the description of the dataset which use the ADA-M profiles \cite{woolley2018responsible} and the license attached to it. Uploading this information also creates the necessary information associated to the dataset within a smart contract in the blockchain network. This sequence of interactions is presented in the simplified diagram displayed in Fig \ref{Sharing} (the complete sequence diagram is shown in Figure \ref{sequenceForSharing} (See appendix).  
\begin{figure}[!h] 
    \centerline{
    \includegraphics[clip=true, scale = 0.35]{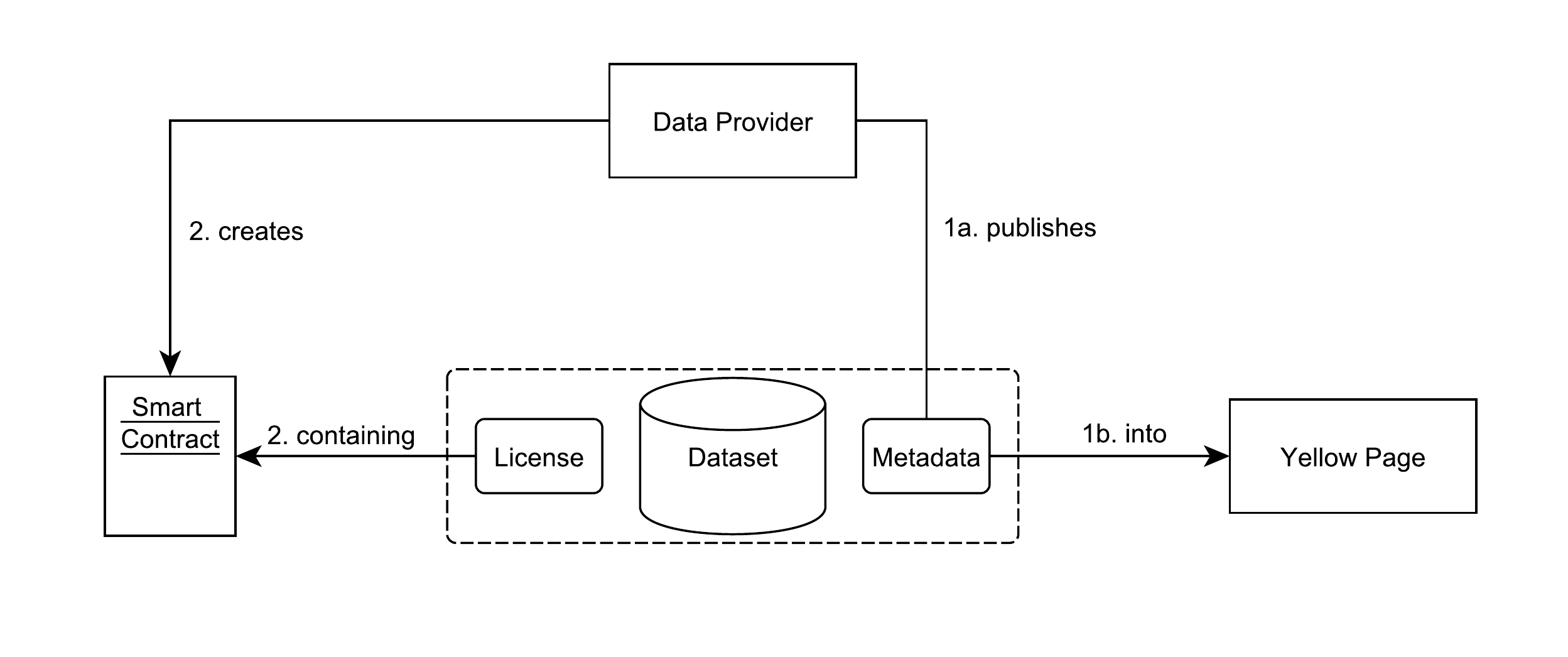}}
    \caption{Sharing a dataset (simplified diagram)}
    \label{Sharing}
\end{figure}


\subsection{Reusing a dataset}
The second scenario covers the case in which a data requester wants to reuse a shared dataset. This happens in two steps. The first is to access the dataset and the second is the monitoring of compliance with the licensing terms of the dataset by the data requester.
\subsubsection{Accessing the dataset}
The first step in reusing a dataset is accessing it, which is illustrated in Fig. \ref{Reusing}. A complete sequence diagram is shown in Fig. \ref{ReusingSequence} (See appendix)


\begin{figure}[!ht] 
    \centerline{
    \includegraphics[clip=true, scale = 0.35]{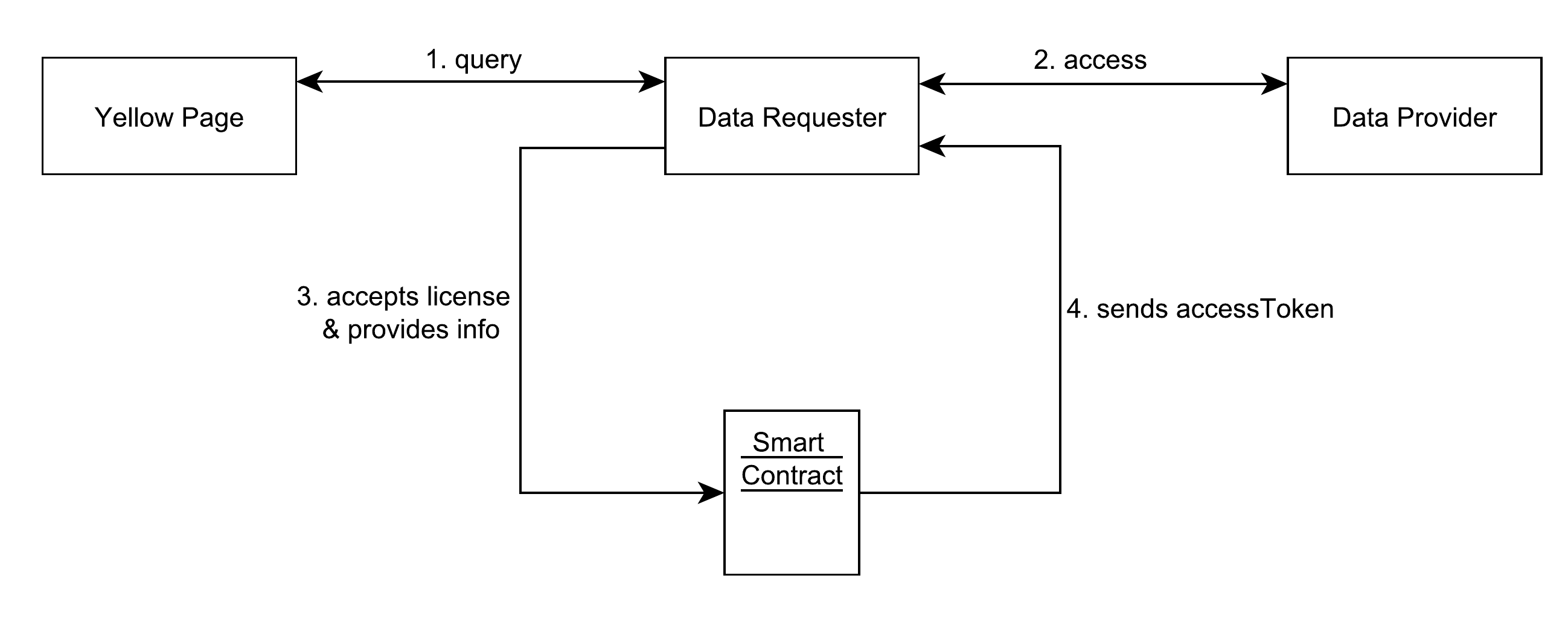}}
    \caption{Accessing a shared dataset (simplified diagram) }
    \label{Reusing}
\end{figure}
\paragraph{Query} Data requesters, who seek to reuse a dataset in the context of their own research, query the yellow page. For each dataset matching the query, the yellow page provides a list containing: the dataset identifier, its meta-data, the type of license attached to it and the data provider's name. Based on this list, the data requester decides which database to reuse. Further negotiations on the data can be set up at this stage as it has been discussed in \cite{dubovitskaya2016multiagent}.

\paragraph{Request} The data requester sends to the data provider a request for access including the dataset's identifier, as well as the purpose for which they want to reuse it. In case the dataset contains data related to data subjects, the purpose indicated by the data requester is checked against the initial purpose these data were collected for. In case both purposes are compatible, the data provider grants access to the data requester by sending them the address of the smart contract associated with the dataset in question. In case the dataset does not contain data related to data subjects, it is still required to specify the purpose of reuse for obtaining the data although is not necessary to verify the compatibility of the purposes. 

\paragraph{Access} 
Once the data requester has been granted access to the dataset, they access the smart contract associated to it. The data requester must agree with the license attached to the dataset.
\paragraph{Accepting licensing terms} Thereafter, the data requester must provide information to the smart contract regarding their use of the dataset. The information to be provided is the same as what would be required in case a data subject exercises their right to access.
\paragraph{Download token} After access has been granted, the smart contract provides the data requester with the link to the repository where the dataset is stored, as well as a token allowing for the download of the dataset. Note that there is only one smart contract associated which each dataset in the yellow page. As such, each data requester willing to use a specific dataset is transacting with the same smart contract. It can identify which data requester it is transacting with thanks to the unique contracted identifier of the data requester. 

\subsubsection{Monitoring compliance with the dataset’s license}

\begin{figure}[!h] 
    \centerline{
    \includegraphics[clip=true, scale = 0.3]{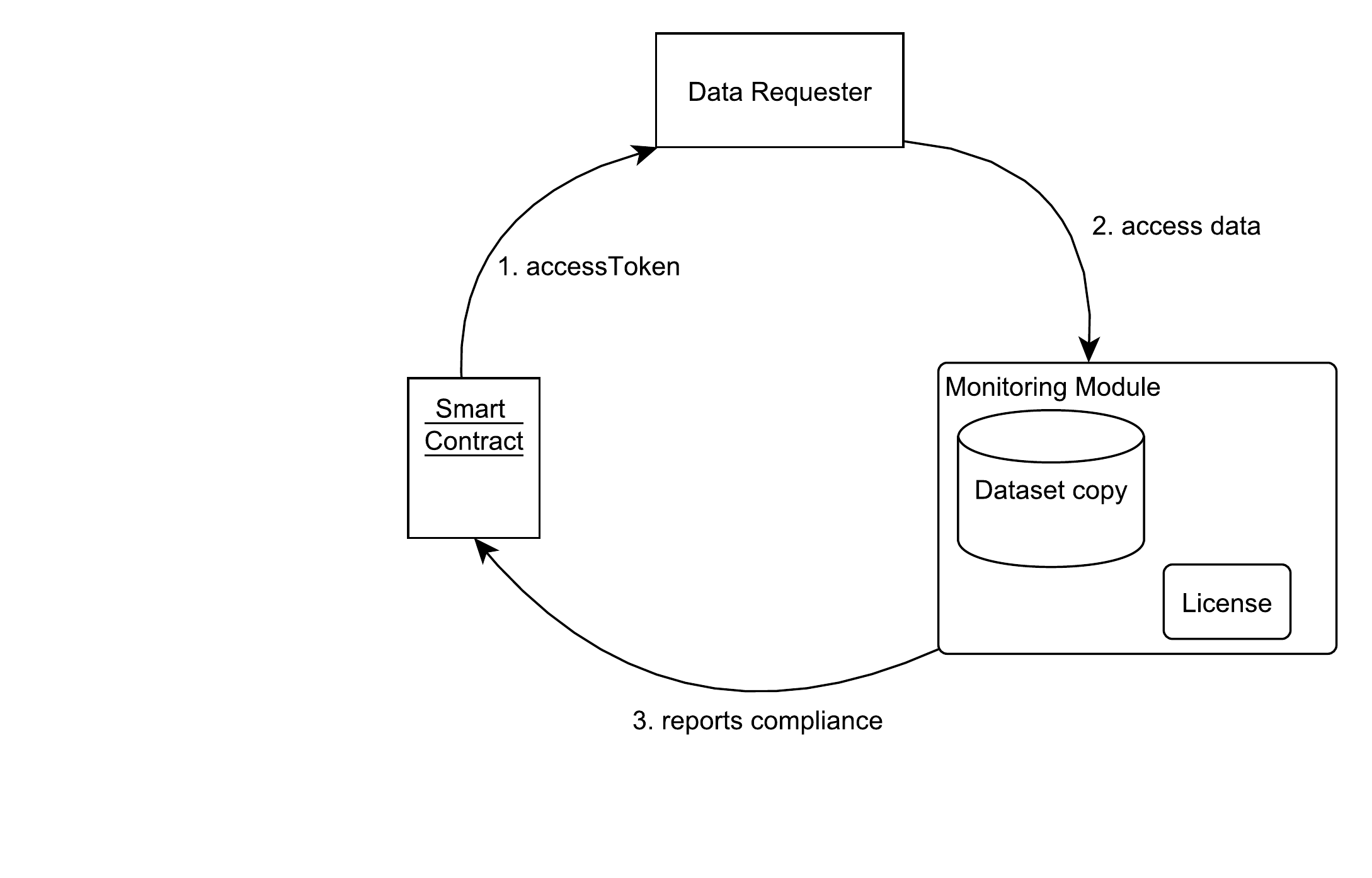}}
    \caption{Monitoring compliance with the licensing terms}
    \label{Compliance}
\end{figure}

The second step in reusing a dataset is the monitoring of compliance with its license by the data requester. This is illustrated in Fig. \ref{Compliance} and described in detail in Fig. \ref{ReusingSequence} (See appendix).
\paragraph{Access token} To access the dataset, the data requester uses the link to its repository as well as the download token. However, they cannot download the dataset alone. For monitoring purposes, it is embodied in an executable, which is encrypted. In order to be able to read it, the data requester needs an access token that they ask to the smart contract.
\paragraph{Access data} Once the data requester has the access token, they can decrypt the executable. It contains a copy of the dataset (in the form of a .cvs file), the machine-readable layer of the license attached to it, as well as a monitoring module. That module is in charge of monitoring what happens with the data, mainly whether it is used in compliance with its license as well as to update the data in case it is needed, e.g. in case a data subject exercises their right to rectification or erasure. The first function is fulfilled by its License Monitoring sub-module while the second one is performed by its Update sub-module. The executable also includes the contracted identifier of the data requester to transact with the smart contract. 
\paragraph{Reports compliance} The License-Monitoring sub-module continuously checks whether the actions performed by the data requester on the data comply with the licensing terms and records these actions in a file which constitutes a log of events. Based on ccREL \cite{CCREL2008} licenses, the report on compliance focuses on the actions of users while using the dataset. In particular we deal with the permissions, prohibitions and requirements of the licence. For example, if the licence states that derivative work should be shared with the same license (share-a-like), then we can monitor that any additional datasets resulting from an original dataset is also shared with the same license terms as the original one.
Periodically, the Update sub-module updates the smart contract as to whether the data requester has complied or not with the licensing terms. It also sends to the smart contract the reference identifier of the log of events as well as a hash of the log (such that it is easy to detect if the log has been modified). A hash is a one-way cryptographic function. As such, it is impossible to retrieve the original input from its "hashed value". Furthermore, the hashed value changes completely when modifying only one character in the original input. Note that the License-Monitoring module never stops checking compliance with the licensing terms and that, as such, it runs in parallel with the Update sub-module.\\
A more flexible way of defining this module is to create a publish-subscribe model and let the user show that they comply with the rules by periodically sending transactions to the contract. In this way we are not fully checking licence agreements, but we have continuous checkpoints and public commitment of the data requester towards licensing terms. 
\paragraph{Access token} The access token is used by the data requester to obtain access to the dataset. The token is only valid during a period of duration T. When that period ends, the data requester asks to the smart contract to renew their access token. The smart contract then verifies whether the data requester has complied with the licensing terms during the last period of time. If so, the contract re-news the access token to the data requester. However, in case the licensing terms were not complied with, the access token is not renewed, meaning that the data requester cannot access the dataset any more. 
\paragraph{Replication} The log of events of the data requester are periodically replicated in the network. At the end of each period of duration T, the Update sub-module sends a copy of the log of events to the replication module, together with the data requester's identifier and the time at which the file has been sent. A new log of events is created, which will also be replicated at the end of the next period of time.
The replication module replicates the log of events on randomly chosen nodes of the blockchain network and associates them with a reference identifier. It is important to note that the replicas are not stored on the blockchain but on the computers of some of the nodes that are part of the blockchain network. The replication module keeps a mapping between the executables, the replicas' reference identifiers and the nodes on which the replicas are stored, making it thereby possible to retrieve replicas in case the log of events does not exist any more. After updating the map, the replication module confirms to the monitoring module that the log of events has been replicated and sends to it the reference identifier of the log. Thereafter the smart contract is updated by the update sub-module.
\paragraph{Checking compliance} A data provider can check whether data requesters have complied with the license terms attached to a shared dataset by accessing the events in the smart contract associated with that dataset.


\begin{figure}[!h] 
    \centerline{
    \includegraphics[clip=true, scale = 0.25]{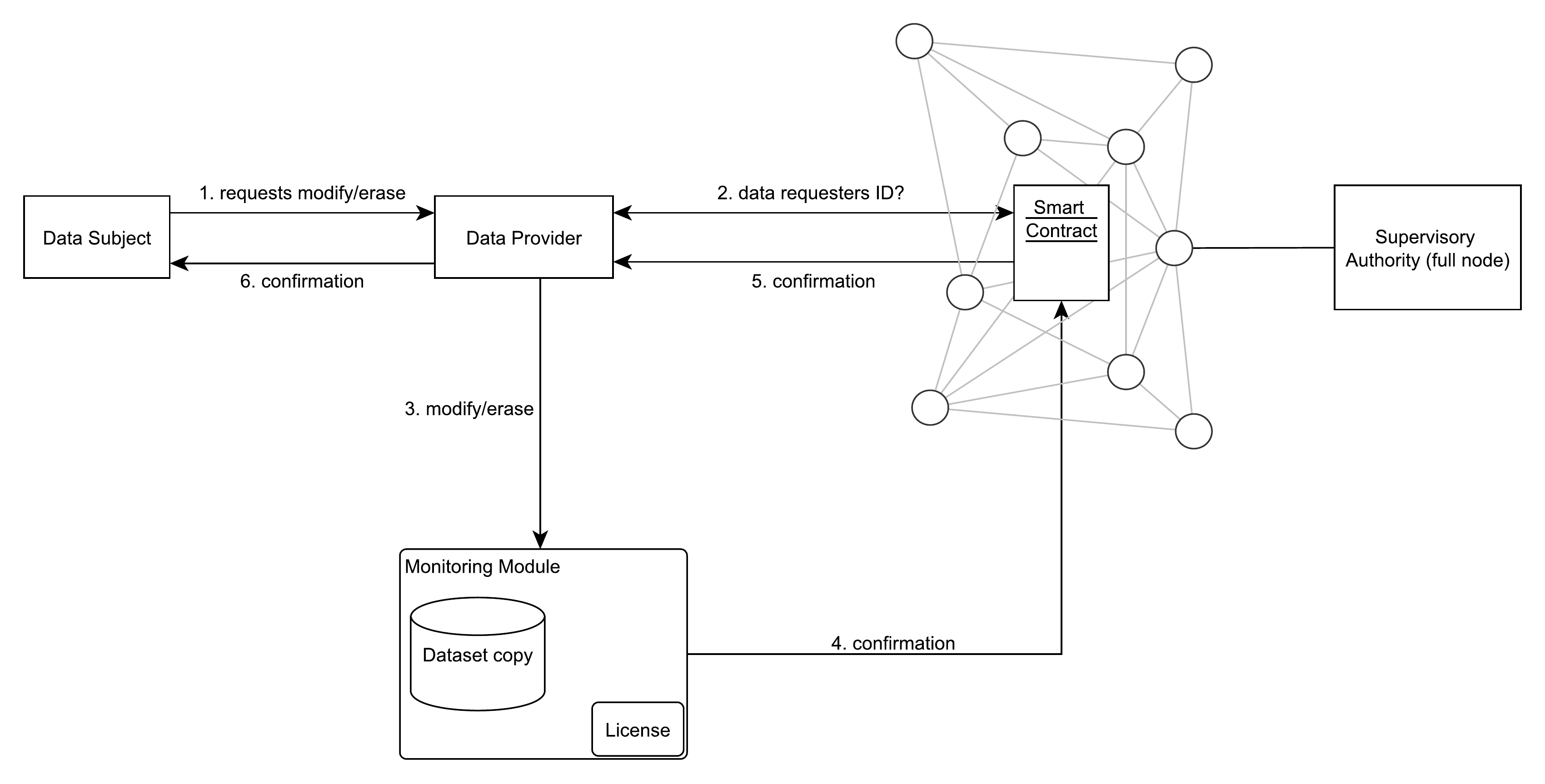}}
    \caption{Complying with GDPR's rights to access, rectification and erasure (simplified diagram)}
    \label{GDPRsimple}
\end{figure}

\subsection{GDPR compliance}
The third scenario our solution covers is the case in which a data subject exercises their right to access, erasure or rectification. How LUCE deals with this scenario is explained below. Fig. \ref{GDPRsimple} displays a simplified diagram of how LUCE deals with the rights to rectification and to erasure (the complete sequence diagram is described in Fig. \ref{GDPR} in the appendix). 

\subsubsection{Right to access} 
In concordance with the GDPR, the data subject has the right to request information from the data provider as to how their data is used, by whom and so forth. Given that the data provider holds a mapping between the data subject and the dataset in which their data record is included, the data provider queries the smart contract associated with the dataset in question and reads it. The smart contract provides the data provider with the information they had stored in it as well as with the information the data requesters that reused the dataset have provided to it. The data provider then sends all this information to the data subject. This must be done for each dataset including data records of the data subject.

\subsubsection{Right to erasure and right to rectification}
\paragraph{Request rectification/erase} In case the data subject requests the data provider to either rectify or erase their data from the dataset, it must be done not only in the dataset of the data provider but also in the copies of the dataset that data requesters are using. The data provider retrieves the anonymized ID of the data subject from the mapping they are keeping between both types of information. The rectification and erasure of a data subject’s records is taken care of by a module called the modification module. The data provider thus sends to the modification module the anonymized ID of the data subject as well as to whether it should erase their record or modify them. 
\paragraph{Data requesters’ ID} The modification module then requests from the smart contract the identifiers of its data requesters. 
\paragraph{Rectify/erase} Once it received this information, the modification module contacts the monitoring modules of the executables including the copies of the dataset in question and asks them to erase or modify the data record corresponding to the anonymized identifier of the data subject. The update sub-module of each of these executables erases or modifies the data record in question from its own copy of the dataset. 
\paragraph{Confirmation} The monitoring module confirms the change to the smart contract associated with the dataset by sending the identifier of the transaction related to the erasure or modification. The modification module then reads that information from the smart contract. It then confirms to the data provider that the rectification or erasure has actually been performed and sends the identifier of that transaction to the data provider who, in turn sends it to the data subject before potentially erasing all information related to him or her. In case the data subject needs the proof that their data have been erased or rectified, they posses the identifier of that transaction and can therefore retrieve it from the blockchain, via the supervisory authority or the data provider.
\subsubsection{Supervisory authority}
Our solution also helps the supervisory authority to check whether the GDPR has been complied with. It indeed is a full node in the blockchain peer-to-peer network, meaning that it can check all the transactions happening via the blockchain platform. Furthermore, it also has access to the replication module.

\section{Prototype of the smart contract} 
\begin{figure*}[!h] 
    \centerline{
   \includegraphics[clip=true, scale = 0.3]{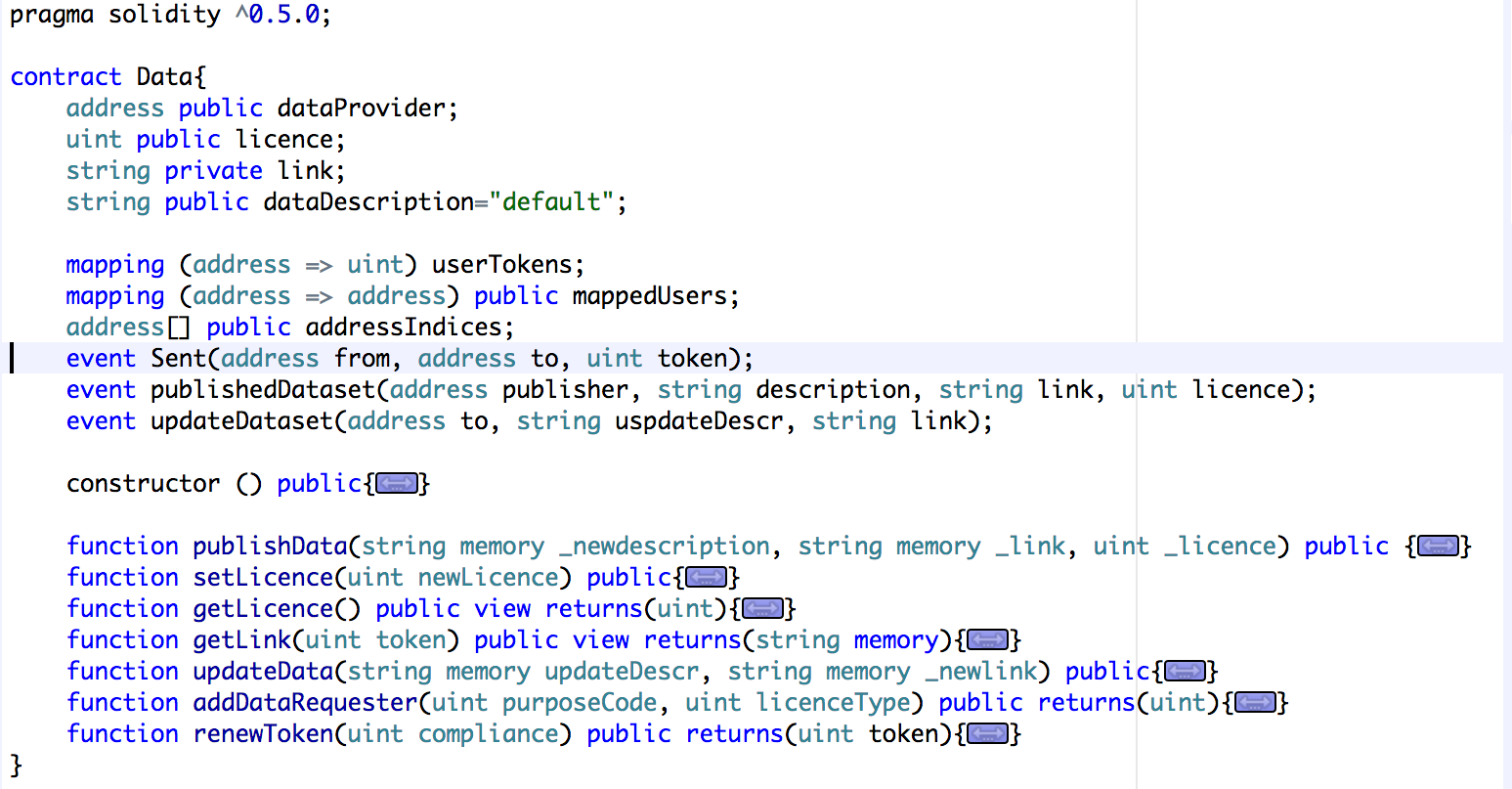}
   }
    \caption{Smart Contracts defined in LUCE}
    \label{Contract}
\end{figure*}
Three possible scenarios were considered regarding the monitoring of the licensing terms compliance. In the first scenario, the smart contract monitors all behaviours. However, this solution would be computationally expensive as the smart contract would be very complex and all  actions performed on the data would be stored on the blockchain. In the second scenario, the smart contract only tracks who has the data but not what happens to them. Even though this is probably the simplest solution, as most of the monitoring would happen off-chain, it does not record enough information to be able to fully inspect what happens to the data. The third scenario involves a smart contract relying on built-in behaviour monitoring. As such, monitoring of licensing terms happens in two stages: some on-chain and some off-chain (the latter can be viewed as sensing). We implemented the third scenario because it provides a good trade-off between complexity and knowledge of the data usage. The only information stored in the blockchain is the one considered as essential. That is, by whom the dataset is used, whether the licensing terms have been complied with and the information that the data provider could be required to provide in accordance with the GDPR. It is however possible to add more details about the actions performed on the data thanks to the replication module.  
\\
We used Solidity to prototype a smart contract for our blockchain platform. Solidity is an event-based programming language for defining smart-contracts in several blockchain networks, amongst which the Ethereum network. It is possible to write and execute a Solidity contract via a Remix, a web-based IDE for Solidity (https://remix.ethereum.org). Figure \ref{Contract} shows our first prototype of the smart contract for data sharing.

 
The shown contract reacts by creating a new event \textit{publishedDataset} when a data publisher decides to share a dataset. When this happens the data provider describes the dataset (\textit{description}), specifies the link to the data (\textit{link}) and the license (\textit{license}). 
Data requesters can then start using a dataset by calling the \textit{addDataRequester} function. In order to call this function, they would have first to read the license terms from the Data contract and indicate the same license in the request. For simplicity, here we have assumed that the license terms will be encoded into a number (e.g. 1001) which, would summarize the actions that can be performed on the data (e.g. sharing is allowed, no re-use, no sending, attribution is needed). This was done in order to keep computation costs for the transaction low, this is due tot he fact that string to string comparison is more expensive than encoding the licence into a number.  \\
Tokens to access data must be renewed via the \textit{renewToken} function. Here the idea is that the monitoring module shown in Fig. \ref{Compliance} will initiate the token renewal on behalf of the dataRequester. The same monitoring module will be able to record and summarize the actions performed on the data. If such actions comply with the license, then the token will always be renewed, otherwise the token is revoked and so is the access to the data.\\
We are working towards fully implementing the LUCE platform and towards testing the performance of the platforms with an increasing number of these smart contracts.

\section{Conclusion and future prospects}
\addcontentsline{toc}{section}{Conclusion and future prospects}
In this paper, we presented LUCE, an architecture of a blockchain-based solution for automatic data management of licensing terms and accountability in a GDPR compliant manner. LUCE enhances the possibilities in terms of data sharing and reuse and makes it easier for researchers to track what is done with their dataset once it has been shared. Moreover, LUCE enables the enforcement of licensing terms and provides a solution for complying with the GDRP's right to access, rectification and erasure. 
This work is however our first step in the development of our solution and some of its components need to be further developed or reflected upon.\\
\textbf{License complying monitoring} - How the License-Monitoring sub-module should check compliance with the licensing terms needs to be further developed. We can check what is done with the data once they have been processed and the data requester becomes a data provider as well. However currently, do not check if the data are shared in another form (i.e by exporting the data into another format). There are also softer ways to monitor re-use, i.e by requesting that the data requester periodically confirms in the blockchain network that is complying with the terms. This means that, should data requesters be found or reported to be in breach of these terms, they may be held accountable.\\
\textbf{Check consent compliance} - In the future, we will further investigate how to be in compliance with the consent given by data subjects. The assumption here is that the license matches the consent of data subjects, which is true in most research datasets, however, there is a need for personalization of the individual consent of the data subjects which is also in line with the GDPR laws. A starting point is the Consent codes developed by the Global Alliance for Genomic and Health \cite{ConsentCodes, ga4gh} to systematically record data usage conditions based on consent, that can be found in the datasets of the main public genome repositories. \\
\textbf{Allow for data integration} - Another point that LUCE can easily address in the future is that datasets may be merged when reused. It actually occurs in practice and is even one of the goals of data sharing and reuse. \\
\textbf{Evaluation} - Our solution also needs to be evaluated and its evaluation should be twofold. Firstly, it should be evaluated in terms of technical feasibility and scalability. Secondly, our solution will eventually help enhance data sharing and reuse only when researchers actually use it. Therefore, some behavioural evaluations are required to investigate whether scientists would be willing to make use of it and what features should be modified, removed or added for them to do so in the end.

\section{Author Contribution}
The idea was generated by Prof. Michel Dumontier, Andine Havelange and Dr. Visara Urovi proposed the LUCE architecture. Furthermore, Dr. Visara Urovi prototyped and developed the LUCE framework. A series of collaborative meetings were set with legal experts (Prof. David Townend and Birgit Wouters) and economics experts (Dr. Jona Linde and Prof. Arno Riedl) who helped shaping the idea from the legal and economics perspective.
The paper was written by Andine Havelange and Dr.Visara Urovi.
Birgit Wouters, Dr. Jona Linde and Prof. Arno Riedl commented on the document.
All authors have read and approved of the paper
\bibliographystyle{plain} 
\bibliography{BlockchainPaperV2}

\section{Appendix}
\begin{figure*}[!b] 
    \centerline{
   \includegraphics[trim = 0.8cm 0cm 0cm 0cm, clip=true, scale = 0.45]{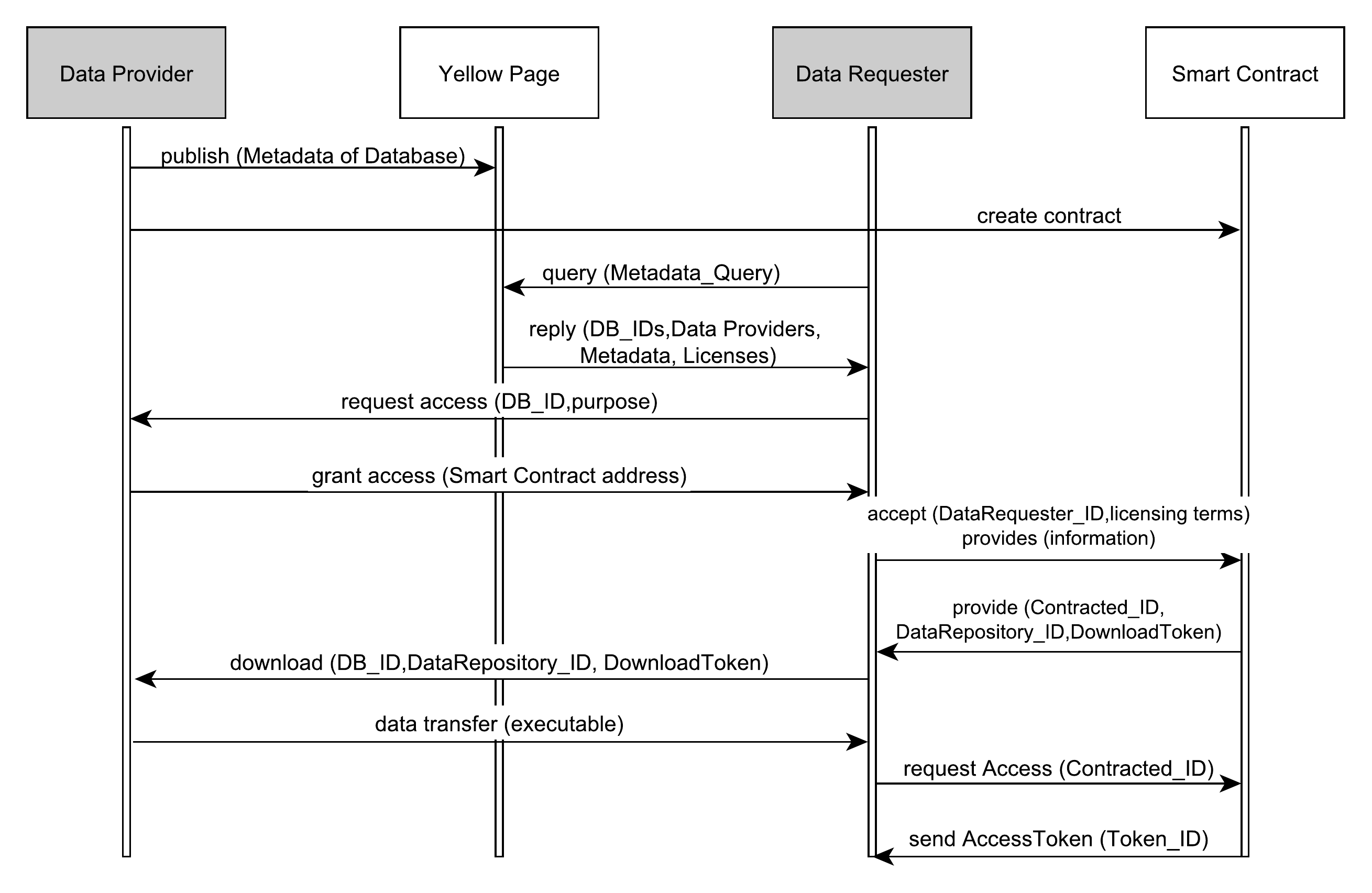}
   }
    \caption{Sharing a dataset: sequence diagram}
    \label{sequenceForSharing}
\end{figure*}

\begin{figure*}[!b] 
    \centerline{
    \includegraphics[trim = 0.5cm 0cm 0cm 0cm, clip=true, scale = 0.6]{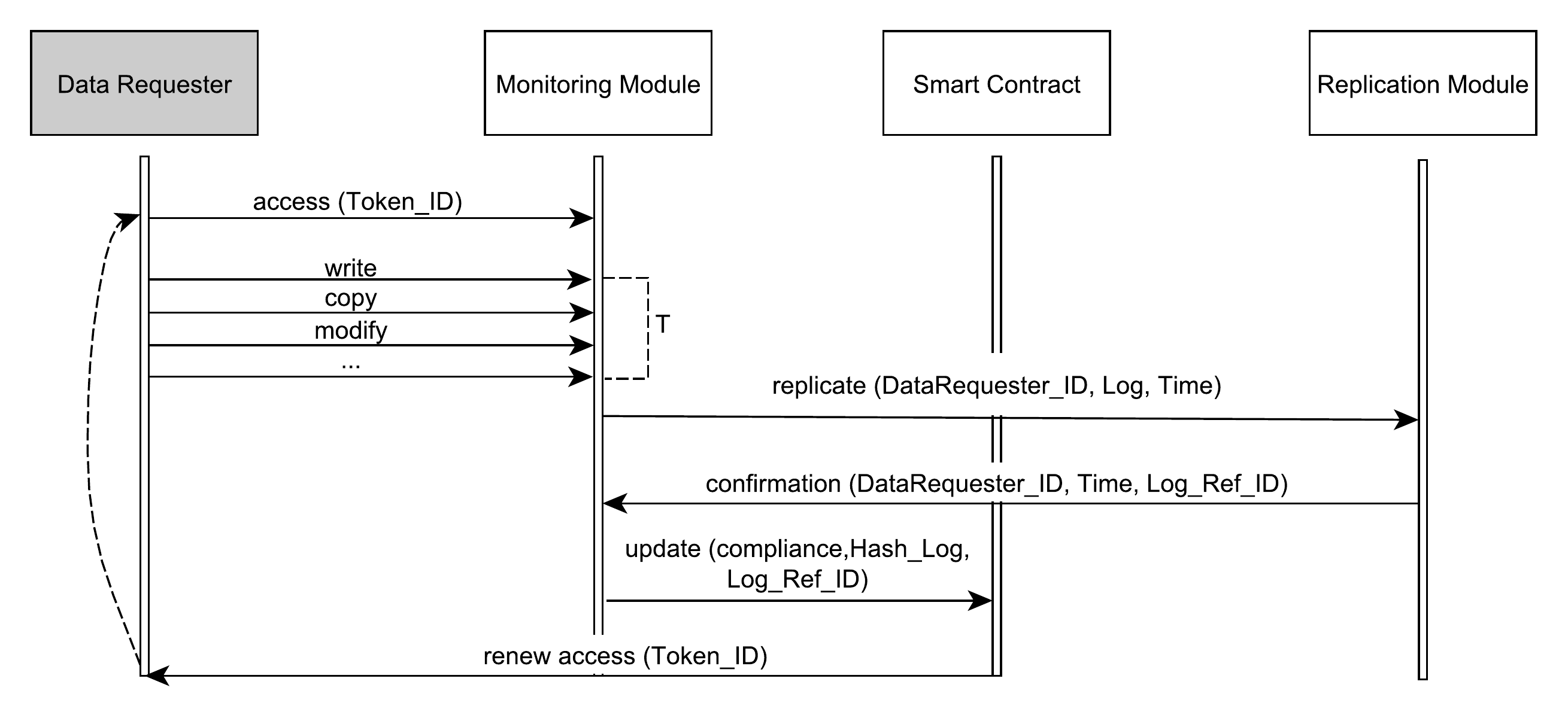}}
    \caption{Reusing a dataset: sequence diagram}
    \label{ReusingSequence}
\end{figure*}

\begin{figure*}[!b] 
    \centerline{
    \includegraphics[trim = 0.5cm 0cm 0cm 0cm, clip=true, scale = 0.5]{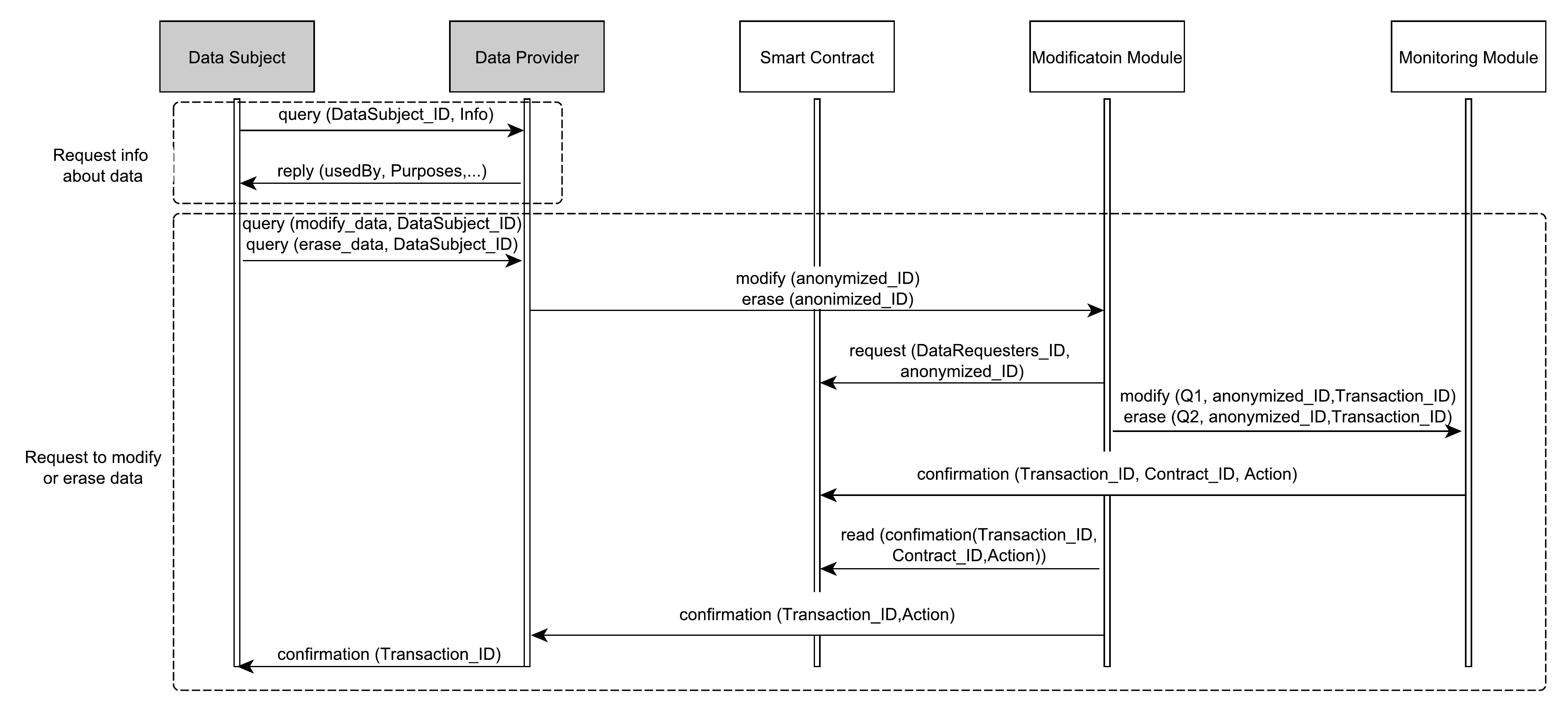}}
    \caption{Complying with GDPR's rights to access, rectification and erasure: sequence diagram}
    \label{GDPR}
\end{figure*}

The following figures show respectively the detailed sequence diagrams for: (i) Sharing a dataset Fig:\ref{sequenceForSharing};
(ii) Re-using a dataset Fig:\ref{ReusingSequence} and (iii) Complying with GDPR Fig:\ref{GDPR}.

\end{document}